\documentstyle[eqsecnum,aps]{revtex}
\input epsf.sty
\begin{document}
%

\title{Mixing of the $f_0$ and $a_0$ scalar mesons in
threshold photoproduction}
\author{B.Kerbikov~\cite{byline}}
\address{ITEP, Moscow,
117218, Russia
}
\author{ Frank Tabakin~\cite{byline} }
\address{
Department of Physics \& Astronomy,
University of Pittsburgh,
Pittsburgh, PA  15260
} 
\def\pslash{{ p \mkern-08mu\raise-.1ex\hbox{/} }} 
\def\kslash{{ k \mkern-10mu\raise-.1ex\hbox{/} }}
\def\epsslash{{\varepsilon  \mkern-08mu\raise-.1ex\hbox{/} }}

\def\la{\mathrel{\mathpalette\fun <}}
\def\ga{\mathrel{\mathpalette\fun >}}
\def\fun#1#2{\lower3.6pt\vbox{\baselineskip0pt\lineskip.9pt
\ialign{$\mathsurround=0pt#1\hfil
##\hfil$\crcr#2\crcr\sim\crcr}}}

\newcommand{\be}{\begin{equation}}
\newcommand{\ee}{\end{equation}}

\date{\today}
\maketitle   
\begin{abstract}
We examine the photoproduction of the light scalar mesons $f_0$-$a_0 (980)$
 with emphasis on isospin-violating  
 $f_0$ and $a_0$ mixing due to the mass difference between neutral and charged kaons.
   General forms for the invariant amplitude for scalar meson photoproduction
 are derived,  which yield expressions for the invariant mass distribution
of the final meson pairs.
 A final state interaction form factor is obtained
which incorporates $f_0$ and $a_0$ mesons, $K\bar K$ threshold
effects, and $f_0-a_0$ mixing. This form factor is applied to predict
the effective mass distribution of the $\pi\pi$ and $K\bar K$ pairs
in the vicinity of the $K\bar K$ threshold.  An estimate of the role of
isospin mixing near threshold is provided.
  The potential role of polarized photons and protons is
also discussed.
\end{abstract}
\pacs{PACS: 25.20.Lj, 13.60.Le, 14.40.cs, 24.10.Eq,  24.70.+s  }
\narrowtext
 \section{Introduction}

The $f_0$ and $a_0$ scalar mesons present a well-known puzzle for which several
 interesting, albeit controversial proposals have been made ranging from
quark-antiquark makeup, to four-quark, and to $\bar{K} K$ molecular structure.  Our
goal is not to review these suggestions, but rather to investigate if
 new information on the makeup of these scalar mesons can be gleaned
from threshold $f_0(980)$ and  $a_0(980)$ photoproduction.  Our hope is that
threshold production of $0^+$  $K\bar K$ and two pion states that arise from the decay
 of scalar mesons will shed light on the nature of the $f_0$ and $a_0$ scalar mesons.
  For that purpose,  we examine a potentially
 important final state isospin mixing effect. 

The $f_0$ has been observed in
 a photoproduction experiment at
Fermilab~\cite{Lebrun}.  Measurements of exclusive $f_0$
photo/electroproduction are being performed at
 Jefferson Laboratory~\cite{CEBAF}.
 Extensive theoretical studies of $f_0$ photoproduction have been presented 
in recent papers~\cite{Ji,MOT}. 
Our study differs
from these works in that we assume $f_0$ and  $a_0$ mesons are
produced directly and play the role of doorway states,
 while in~\cite{Ji,MOT} these
mesons enter via final state interaction in   $K\bar K$ and  $\pi \pi$ channels.  The
two descriptions are complementary and further studies are needed
to clarify their interconnection.  Another novel feature of our present
work is the inclusion of possibly important final state  $f_0 - a_0$ mixing.
This mixing is induced by  scalar meson transitions into and out of $K^+ K^-$ and
 $K^0\bar K^0$ intermediate states, which generates
 $f_0 - a_0$ mixing because of the 8 MeV
splitting between the $K^+ K^-$ and
 $K^0\bar K^0$ thresholds.  Thus scalar meson mixing arises from the difference 
between the neutral and charged kaon masses, e.g. from isospin breaking.
This effect is fully included into our treatment.  Calculations presented below rely
on a very restricted number of
parameters and most of which are fixed using recent experimental
data on the $\phi \rightarrow \gamma f_0/a_0$ reaction~\cite{Achasov1,Novo} .

We study the reaction
\be
\gamma + p \to p' + f_0/a_0 \to p'+m_1 m_2 \ ,
\label{reaction}
\ee using the $f_0$ and the $a_0$ as the possible doorway for subsequent
decay meson production.
Here $m_1 m_2$ denotes $\pi\pi$ or $K\bar K$ pairs, or the $\pi\eta$ system.
A thorough theoretical investigation of this reaction should include
at least three essential points:

a) the general structure of the invariant  scalar meson photoproduction 
amplitude,
including expressions for the cross sections and spin observables;

b) effects of the  final state interaction, including $K\bar K$ threshold
phenomena;

c) a dynamical model for the reaction mechanism, e.g., an effective Lagrangian
and/or a set of leading diagrams.
In our paper we concentrate on points (a) and (b), which are more universal and less
 model dependent than point (c) above. 
  The structure of the basic production amplitude and spin
observables presented below do not rely on any explicit reaction
mechanisms (other than the doorway assumption) and are applicable to the
photoproduction of any scalar meson.
 The final state interaction (FSI) form factor we present later
 is constructed in a model independent way based on a theory of resonance mixture.  Only a few
parameters are needed which on one hand are sensitive to the nature of the $f_0/a_0$ mesons, and on the
other can be extracted from recent experimental data on the $\varphi \rightarrow \gamma f_0/a_0$
 reaction\cite{Achasov1,Novo}.
  Concerning the dynamical point (c), the dominant reaction mechanism is expected to depend upon
the kinematical conditions of the explicit  experiment.  For example, at low photon momenta $s$--channel 
resonances might give a substantial contribution~\cite{Tejedor}, while  forward photoproduction at higher
energies might be dominated by 
$t$--channel,  $\rho$ and $\omega$ meson exchanges~\cite{Ji}. An alternate, very promising
 approach making use of chiral Lagrangians was proposed recently~\cite{MOT}. 
 Although we do not propose an explicit reaction mechanism, our driving idea is
complementary to that of  Refs.~\cite{Ji,MOT}. We take the $f_0/a_0$ to be ``doorway states" for the
reaction.  Namely, we assume that $f_0/a_0$ scalar mesons are produced
``directly'' and then propagate under the strong influence of two
nearby $K \bar{K}$ thresholds before
decaying into $K \bar{K},  \pi \pi,$ or $\pi \eta.$ 
This propagator, or FSI form factor,  has a simple form and is applicable to any reaction involving
$f_0/a_0.$

We begin by analyzing the structure of the photoproduction of scalar mesons
 using this $0^+$ 
doorway model,  then
we formulate a description of the final state interaction.  The general behavior
of the cross-section as a function of the invariant mass of the
kaon or pion pair is then generated to gauge the importance of the mixing near
threshold
and for comparison with experiment. Speculations about the role of scalar-meson
nucleon P-waves are presented,   especially as they relate to the possible
observation of spin observables.

\section{Diagrams and cross sections}

Before focusing on $f_0/a_0$ photoproduction in
reaction (~\ref{reaction}),  we comment on possible background effects.
 
The amplitude for two-meson photoproduction $\gamma p \to p'
m_1m_2$ can be   represented symbolically by the series of  graphs
depicted in Fig.~1. 
 Although all diagrams entering into the background term $B_\alpha$~\footnote{
The subscript $\alpha$ denotes the $ \pi \pi,$ or $ K \bar{K}$ channels.} 
(Fig.~2) might be
explicitly calculated (some of them were calculated in~\cite{Ji}),
 our main point is that $B_\alpha$ is 
a smooth function of the energy in the
vicinity of the $K\bar K$ threshold;  whereas,  dramatic
 energy dependent effects in the
invariant mass distribution of the $m_1 m_2$ pair (see below) are
generated mainly by the first two diagrams in Fig.~1. We neglect the pion loop
diagram shown in Fig.~3 (see~\cite{Ji,MOT})for two reasons.  One reason is that
 near the $K\bar K$
threshold the pion loop by itself, i.e., without adding
 a resonant rescattering vertex, 
 has smooth  energy behavior and  has to be included
into the renormalized vertex of the $f_0$ production. The second reason for
neglecting  pion loops  is based on Weinberg's power counting 
theorem~\cite{Weinberg}, which states that pion loops are suppressed  compared
to the tree diagrams depicted in Fig.~1. 
  According to these arguments the pion loops are of the same order
as higher terms in the effective Lagrangian.   However,
the above reasoning is far from being a rigorous proof of the direct production dominance.
The theoretical and experimental study of this problem is
 at an early stage and we hope that
 comparison of complementary approaches
will advance the investigation.

Consider now the graphs on the top of Fig.~1. 
 We first consider only the $f_0$-meson; 
 $a_0$ production will be
included later.  Then  these diagrams under consideration are described by the following equation

\be
T_{\alpha} = V_{\alpha f} G_f F_f + B_\alpha,
\label{amp1}
\ee
where $F_f$ is the $f_0$ photoproduction amplitude, $G_f$ is the $f_0$ propagator
(also called the FSI form factor), and $V_{\alpha f}$ is the amplitude for $f_0$ decay into the
final meson pair 
channel $\alpha$ ($\alpha \equiv \pi \pi, K \bar{K}$),  $B_\alpha$
denotes the background terms for channel $\alpha$ which we presume to be weakly energy dependent.

 As shown below, the form factor $G_f$ contains dynamical information on the nature
 of the $f_0$ meson.  This form factor also reflects the interplay of the $f_0$ with the nearby $K
\bar{K}$ threshold.   The simple kinematical fact that $K^+ K^-$ and $K^0 \bar{K}^0$ thresholds are split
by 8 MeV induces significant $f_0/a_0$ mixing, as first
 noticed by Achasov et al.~\cite{Achasov2}.
 Taking
account of this $f_0/a_0$ mixing,  Eq.~(\ref{amp1})  is now extended to
 include a sum over intermediate $f_0$ {\it and} $a_0$ states
\be
T_{\alpha} = \sum_{i, k}\ V_{\alpha i} G_{i k} F_k + B_\alpha,
\label{amp2}
\ee
with $ i, k = f_0, a_0 $, i.e., the form factor $G_{i k}$ becomes a $2\times 2$ matrix
in $f_0/a_0$ space. 
 As  pointed out earlier by Stodolsky~\cite{Stodolsky1} measurement of the invariant mass 
distribution of the $(m_1 + m_2)$ system in reaction (~\ref{reaction}) enables one
 to determine the nonorthogonality of
the decaying states; i.e. the probability for $f_0 \leftrightarrow a_0$
transitions.

The cross section for the 
 reaction $\gamma p \to p' m_1 m_2$  (see Fig.~1),  where the meson pair
is in channel $\alpha,$ is
\be
\sigma_\alpha = \frac{m^2}{ (2 \pi)^3} \frac{1}{\lambda^{1/2} (s,0,m^2)} \int dE_{m_1}\  
dE_{ m_2}\ \frac{d\Omega_{m_1}}{4 \pi}\ \frac{d\varphi_{m_2}}{2 \pi}\  |T_\alpha|^2,
\label{cross1}
\ee 
\narrowtext
\noindent where $\lambda(x,y,z) \equiv (x-y-z)^2-4yz,\  s=(k+p)^2=(p'+q_{m_1}+q_{m_2})^2,\ 
 m$ is the proton mass, $\Omega_{m_1}$ is the angle of decay meson 1
 with respect to the incident photon beam 
$\vec k$ and $\Omega_{m_2} = (\cos \theta_{m_1 m_2}, \varphi_{m_2})$
is the angle of decay meson 2 with respect to $\Omega_{m_1}$.
The decay mesons are coplanar as described by $\varphi_{m_2}.$

 To observe the $f_0/a_0$
meson and to investigate its properties use is made of the double
differential cross section  
\be
\frac{d\sigma_\alpha}{dt\ dm_{x}}= \frac{m^2}{ (2 \pi)^3}\ 
\frac{\sqrt{\lambda(m^2_x, m^2_{m_1},m^2_{m_2})}}
{4\ m_x\ \lambda(s,0,m^2)}\ \frac{1}{4 \pi}\ \int d\Omega^*\ |T_\alpha|^2 ,
\label{cross2}
 \ee or of the invariant mass $m_x$ distribution
\be
\frac{d\sigma_\alpha}{dm_{x}}=  \frac{m^2}{ (2 \pi)^3}\ 
 \frac{ \sqrt{\lambda(s, m^2_{x},m^2)\   
 \lambda(m^2_x, m^2_{m_1},m^2_{m_2}) }}
{4 \ s\ m_x\ (s - m^2)} \frac{1}{2} \int^{+1}_{-1} d \cos \theta\ 
\frac{1}{4 \pi}\ \int \  d\Omega^*\  |T_\alpha|^2 ,
\label{cross3}
\ee where $t=(p-p')^2,
m_{x}=s^{1/2}_1, s_1 =(p_{m_1}+p_{m_2})^2,$ and $\Omega^*$ is the angle between the
relative momentum of the two decay mesons in their c.m. system.
The angle $\theta$ refers to the scalar meson production angle.
  Proton spinors are
normalized according to $\bar{u} u = 1.$ 
 We will use this last expression to display the cross-section as a
function of the invariant mass $m_x$, once we have described the amplitude $T_\alpha$
 of Eq.~(\ref{amp2}).

The initial and final helicity summations are suppressed above  
 in $|T_\alpha|^2.$  For unpolarized beam, target and recoil proton
experiments,  it is therefore understood that one
sums over final and averages over initial helicities.
However, if one has a polarized beam or polarized target,  the associated cross sections
can be expressed,  using the bold assumption that $B_\alpha$ interference can be neglected, by:
\be
|T_\alpha|^2 = V_{\alpha i} G_{i k}\  Tr[(F_k \rho_I F^\dagger_{k'})]\ G^*_{i' k'}V^*_{\alpha i'}\ ,
\label{spin1}
\ee where $\rho_I$ is the density matrix which describes the
spin state of the initial beam and target. 
Since the decay $ V_{\alpha i}$ and the propagator $G_{i k}$ are both independent of
helicities,  the inner term above ${\cal S}_{k k'} = Tr[(F_k \rho_I F^\dagger_{k'})]$ includes a trace over the
helicity quantum numbers in the usual way.\footnote{ Note that
 ${\cal S}_{ k k'}^*  = {\cal S}_{k'k}$
which assures real observables and that ${\cal S}$ is Hermitian in the $f_0/a_0$ channel space,
which should not be confused with the Hermiticity of the density matrix in spin-space.}\
 The function ${\cal S}_{k k'}$ is a coupled-channels version of the usual ensemble average expression. 
For a polarized recoil proton experiment, ${\cal S}_{k k'}$ generalizes to
 ${\cal S}_{k k'} = Tr[\rho_F (F_k \rho_I F^\dagger_{k'})],$  where $\rho_F$ describes the final recoil
polarization measurement.
With this expression,  it is possible to
extract the effect of having a polarized beam and/or target on the cross-section.
It is of interest to see if such measurements,  especially as  meson-nucleon P-waves 
appear, can
shed light on the structure of the scalar mesons.

 We are now ready to examine the structure of scalar meson photoproduction in order  to specify the
scalar meson production amplitude $F.$

\section{Invariant Amplitudes and Spin Observables for the Scalar
Meson}

The spin structure of the scalar meson
photoproduction amplitude $F$ and the corresponding spin observables are discussed in this section.
Again $k$, $p$ and $p'$  denote the 4-momenta of the photon and
 the initial and final  protons, respectively;  while the scalar meson's 
4-momentum is denoted by $q.$ The
 decomposition of the {\it scalar} meson photoproduction amplitude as invariant amplitudes
 has the following (CGLN~\cite{CGLN})form
\be
F=\bar u (p') \sum^4_{j=1} M_j A_j(s,t,u) u(p).
\label{CGLN}
\ee
We follow the $\bar{u} u = 1$ normalization conditions of~\cite{BD}. For the scalar meson, 
the four invariant amplitudes $M_j$ are 
$$
M_1=- \epsslash \kslash,
$$
$$
M_2=2  (\varepsilon p)(kp')- 2 (\varepsilon p')(kp),
$$
\be
M_3=   \epsslash\ (kp) - \kslash\ (\varepsilon p),
\label{invM}
\ee
$$
M_4= \epsslash \ (kp')-\kslash\  (\varepsilon p') ,
$$
where $\varepsilon_\mu$ is the photon polarization 4-vector. The four amplitudes $M_i$
differ from the $0^-$ meson photoproduction only in the omission
of the $\gamma_5$ pseudoscalar factor.
 
The
amplitude $F$ can be expressed in terms of the two--component spinors
$\chi$. For this  purpose we write
\be
F=\sqrt{\frac{\omega(p)\omega(p') }{4m^2}} <\chi(p')|R|\chi(p)>,
\label{inv1}
\ee
where $\omega(p)=E_p+m$. Substitution of the invariant amplitudes Eq.~(\ref{inv1})
into Eq.~(\ref{CGLN}) leads to the following CGLN-type representation for the amplitude $R$
\be
R=i \vec \varepsilon\cdot(\vec \sigma \times \hat k)\  {R_1}+
 (\vec \varepsilon \cdot \vec \sigma )( \vec \sigma \cdot \hat q)\  {R_2}
+i (\vec \varepsilon \cdot \hat q)  (\hat q \times \hat k) \cdot \vec
\sigma\  {R_3}+ (\vec \varepsilon \cdot \hat q)\  R_4 \ ,
\label{inv2} 
\ee where $\hat{k}, \hat{q}$ are unit vectors. 
The four amplitudes $R_j$ are related to the four amplitudes $A_j$
of Eq.~(\ref{CGLN}) via
$$
R_1=-(\sqrt{s}-m) \Big\{ A_1+\frac{1}{2}(\sqrt{s}-m)
A_3+\frac{(kp')}{\sqrt{s}+m} A_4 \Big\}
$$
\be
R_2=\frac{|\vec{ q}\, |}{E_{p'}+m} \Big\{ -(\sqrt{s}-m)A_1+\frac{1}{2}(s-m^2)A_3+(kp') A_4 \Big\}
\label{inv3}
\ee 
$$
R_3=|\vec{ q}\, |^2
(\frac{\sqrt{s}-m}{E_{p'}+m}) \Big\{ (
\sqrt{s}-m)A_2-A_4 \Big\}
$$
$$
R_4=-(\sqrt{s}-m)\ |\vec{q}\, |\  \Big\{ A_2(\sqrt{s}+m)[1-
\frac{\vec k \cdot \vec q}{(E_p+m)(E_{p'}+m)}]+
$$
$$
+A_4[1+(\frac{\sqrt{s}+m}{\sqrt{s}-m})
\frac{\vec k \cdot \vec q}{(E_p+m)(E_{p'}+m)}] \Big\}.
$$ The $(kp')$ factor denotes a
4-vector product.
To recast ~(\ref{inv3}) into a fully relativistic invariant form, the following
 simple kinematical equations may be used
$$
2(\vec k \cdot  \vec q) =t-\mu^2 +\frac{1}{2s} (s-m^2) (s-m^2+\mu^2),
$$
$$
2kp'=m^2-u,
$$
\be
|\vec q\, |=\frac{1}{2\sqrt{s}}\lambda^{1/2}(s,\mu^2, m^2),
\label{inv4}
\ee
$$
E_p=\frac{1}{2\sqrt{s}}(s+m^2), E_{p'}=
\frac{1}{2\sqrt{s}}(s+m^2-\mu^2),
$$
where $\mu$ is the mass of either  $f_0$ or $a_0.$  Note  that an
alternative form of the amplitude~(\ref{inv3}) may be found in Ref.~\cite{Bilenky}.

Note that at the scalar meson production threshold
all of the $R_i$ amplitudes  vanish except for $R_1.$
We conclude that at the scalar meson production threshold the operator $F$
is given by
\be
F=\frac{[\omega(p)\omega(p')]^{1/2}}{2m}\  R_1\  i \vec \varepsilon\cdot(\vec \sigma \times \hat k).
\label{ inv5}
\ee The value of the  coefficient $R_1$ is discussed below in
Sec.~5.

  As meson-nucleon  P-waves turn on,
 the associated $J= \frac{1}{2}^- , J= \frac{3}{2}^- $ amplitudes contribute  
with an initial linear dependence on the momentum $|\vec{q}\ |;$
 hence, it is likely that the terms $R_2$ and $R_4$  will appear along with their operators.  That
unfolding of P-waves has implications concerning which spin observables assume 
nonzero values as the energy rises beyond the threshold
 value of $E_\gamma(lab) = 1.5\ $ GeV.

\section{The Form Factor of the scalar Mesons near the $K\bar{K}$
Threshold}

In this section, we consider the FSI form factors $G_f$(pure $0^+$ propagation)
 and $G_{ik}$(coupled $0^+$ propagation) introduced in (\ref{amp1}) and
(\ref{amp2}).  These form factors describe the propagator of an unstable particle in the
 energy region overlapping
some of the decay thresholds ($K^+K^-$ and $K^0 \bar{K}^0$).  We are  mainly interested in the
coupled-channel form factor
$G_{ik},$ which includes  $f_0-a_0$ mixing, but to simplify the discussion, we  start with the simple
one-channel propagator $G_f,$  which describes the $f_0$-meson and its decay channels.  Later we generalize that
discussion to the coupled-channels $G_{ik}$ case.

There exists an overwhelming number of approaches to describe an unstable
particle
 and particle mixtures. 
 We  follow the general phenomenological approach developed by Stodolsky~\cite{Stodolsky2} and
Kobzarev, Nikolaev and Okun~\cite{Kobzarev}, and in a slightly modified form presented in the book of
Terent'ev~\cite{Terent'ev}.  This approach has  already been applied to the $f_0/a_0$ system in~\cite{Bashinsky}. 
 We now briefly outline the derivation of the basic equations for propagation of
an unstable particle; see~\cite{Stodolsky2,Kobzarev,Terent'ev,Bashinsky} for details.

  First consider $f_0$ and its decay channels ($a_0$ will be incorporated
later) and let us introduce a 
  set of state vectors $\mid i >$ describing the scalar meson.
  One of these states, often denoted as 
 $\mid f >,$ is a discrete state; continuum states are 
also included in the set $\mid i >.$ 
 The discrete state couples to the continuum states 
 and thereby acquires a width
and a shift in mass; e.g.,
  it becomes an unstable state.  Let the whole scalar meson system be
described by the Hamiltonian
$H = H_0 + V$, where
$H_0$ has a multichannel continuous spectrum $H_0
\mid i > = E_i \mid i >$ and a discrete state $H_0 \mid f > = E_f \mid f >.$  The interaction V is responsible
for the transitions between the above channels, so that with V ``turned on,'' $\mid f >$ becomes a resonance. 
Consider the transition amplitude  

\be
 A_{if} (t) \equiv < i \mid f; t >,
\label{time1}
\ee where $\mid f; t > = \exp (-i (H_0 + V) t) \mid f >$.  
``Diagonal" transitions $A_{ff}(t)$ are also included in this definition.
  The amplitude $A_{if} (t)$ satisfies the equation
\be
i \frac{\partial}{\partial t} A_{if} (t) = E_i\  A_{if} (t) + \sum_{j}\  V_{ij}\  A_{jf}  (t),
\label{time2}
\ee where $V_{ij} = < i \mid V \mid j >,$ and the initial condition is $A_{if} (0) =
\delta_{if}$.  The equivalent integral equation reads

\be 
A_{if} (t) = e^{-i E_i t} A_{if} (0) - i\  \sum_{j} \int^{t}_{0} d t'\  V_{ij}\  e^{-i E_i (t-t')} A_{jf}
(t').
\label{time3}
\ee

Next we introduce the propagator for the interacting
scalar meson system in the energy representation

\be
G_{if} (E + i \delta) = \int^{\infty}_{0} d t\ e^{ i (E + i \delta) t} A_{if} (t).
\label{time4}
\ee
Applying a time
integration similar to 
 Eq.~(\ref{time4}) to all of Eq.~(\ref{time3}) yields

\be
G_{if} (E + i \delta) = \frac{i A_{if} (0)}{E-E_i + i \delta} + \sum_{j} V_{ij}\  \frac{G_{jf} (E + i \delta)}{E
- E_i + i \delta}.
\label{green1}
\ee

For $i \neq f$ taking account of the initial condition,  one gets
\be
G_{if} (E + i \delta) = V_{if}\  \frac{G_{ff} (E + i \delta)}{E-E_i
+ i \delta} + \sum_{j \neq f} V_{ij}\  \frac{G_{jf} (E + i \delta)}{E-E_i + i \delta}    \ ,
\label{green2}
\ee  while for $i = f$ the equation  has the form
\be
G_{ff}( E + i \delta) = \frac{i}{E-E_f + i \delta} + 
V_{ff}\  \frac{G_{ff} (E + i \delta)}{E-E_f + i \delta} + \sum_{j \neq f} V_{fj}\  \frac{G_{jf} (E + i
\delta)}{E - E_f + i \delta}     \ .
\label{green3}
\ee
Now we return to ~(\ref{green2}) and assume that $V$ only connects $| f>$ with its decay
channels,  while direct coupling between decay channels is absent,
i.e., $V_{ij} = 0$ in Eq.~(\ref{green2}) (this assumption
may be dropped without changing the results significantly  see Ref.~\cite{Stodolsky2}). 
   Then we take $G_{if}$
given by the left hand side of Eq.~(\ref{green2}) with $V_{ij}=0 $ on the right hand side,
  change the index $i$  in $G_{if}$ into $j$ and substitute this
 $G_{jf}$ into ~(\ref{green3}).  Thus we obtain
\be
\Big( E - E_f - V_{ff} -\sum_{j \neq f} V_{fj}\  \frac{1}{E-E_j + i \delta} V_{jf}
 + i \delta \Big) G_{ff} (E + i \delta) = i .
\label{green4}
\ee  This is the result for a pure $f_0$ meson case. The physical meaning of this
propagator is that there is a shift in energy of the
interacting meson due to a self interaction,  plus a complex contribution from
transitions to intermediate continuum states. Some of the flux into the
intermediate state returns to the discrete state $| f>,$ some flows away, thereby
creating a shift in the width as well as the real part of the energy.
Note, we often designate the interacting propagator $G_{ff}$  simply as $G_f.$
We now examine the role of the scalar mesons coupling
to pion and kaon pairs in the intermediate states.

Some remarks are now in order.  Since we are interested in 
 $K \bar{K}$ channels with 
 thresholds close to the $f_0$ mass  the non-relativistic
 derivation and in particular
nonrelativistic kinematics used above are justified; the relativistic generalization is straightforward. 
 Keeping in mind that
$\pi^+ \pi^-$ and $\pi^0 \pi^{0}$ thresholds are far away from the $f_0$ position, 
 we rewrite Eq.~(\ref{green4})
in terms of ``renormalized'' eigenvalues instead of the ``bare'' ones.  
By that we mean that $E_f $ is redefined to include both  
the self-energy contribution $V_{ff}$ and the part of the sum in Eq.~(\ref{green4}) 
that arises from  channels
other than the
$K \bar{K}$ channels. Hence, effects of the $\pi^+ \pi^-$ and $\pi^0 \pi^{0}$
are incorporated by redefining or renormalizing $E_f.$
 In the vicinity of the $K \bar{K}$ threshold, the
terms arising from intermediate pion pairs are almost energy independent.
  They shift $E_f$ by the following real and imaginary parts
\be
\sum_{\pi \pi} V_{f \pi \pi} \frac{1}{E-E_{\pi \pi } + i \delta} V_{\pi \pi f} = M_f -i \frac{\Gamma_f}{2}.
\label{isospin}
\ee The width $\Gamma_f$ is thus generated by the imaginary
part of the transitions to the pion pair continuum.
  For the  $a_0$-meson propagator analogous real
and imaginary energy shifts arise from  $ \eta \pi$ intermediate states.
The sum in~(\ref{isospin}) implies both a sum over different channels ($\pi^+ \pi^-$
and $\pi^0 \pi^0$) and integration over the energies $E_{\pi \pi}$ in each channel.
  For large values of $E_{\pi \pi}$ the integral
diverges and standard renormalization has to be performed. 
 In the simplified approach presented here we may
even argue that the integral is convergent (or cut off) due to
the matrix elements $V_{f \pi \pi}$, $V_{\pi \pi f}$. 
 We  keep the same notation $E_f$ for the real part of
 the $f_0$ energy renormalized in the above
sense; i.e., after
$V_{ff}$ and $M_f$ are absorbed into it.  Then the $f_0$ propagator takes the following simple form

\be
G_f (E + i \delta) = \frac{i}{E - E_f + i \frac{\Gamma f}{2} - 
\sum_{K \bar{K}} V_{f K \bar{K}} \frac{1}{E - E_{K \bar{K}} + i \delta} 
V_{K \bar{K} f}}\ \  ,
\label{green5}
\ee
where summation is now only
 over $K^+ K^-$ and $K^0 \bar{K}^0$ channels and the
 sum also implies integration over energy.
Being interested in the $K \bar{K}$ near threshold energy region,
 one cannot neglect the splitting between $K^+
K^-$ and $K^0 \bar{K}^0$ thresholds equal to 2 ($m_{K^0} - m_{K^\pm} )$ = 8 MeV. This mass difference induces
isospin violating $f_0 - a_0$ mixing. 
 Such ``kinematical'' isospin violation was carefully studied earlier using 
effective range theory~\cite{Ross}.  For the $f_0-a_0$ system the effect
 was probably first pointed out in~\cite{Achasov2},  where it
was shown that mixing is enhanced, i.e.,  is of the order of 
$ [ (m_{K^0} - m_{K^\pm}]/m_{K^0}]^{\frac{1}{2}},$  instead of
$[ (m_{K^0} - m_{K^\pm}]/m_{K^0}]$  as might be naively expected.
 As we shall see, this enhancement also follows in our approach and motivates
us to include this effect in scalar meson production.
Isospin violating $f_0 - a_0$ mixing has also been studied earlier
by T.~Barnes.~\cite{Barnes}

We will not repeat the derivation including the $a_0$ meson 
 since it proceeds along the same lines. 
 The propagator now becomes a $2 \times 2$ matrix in the
$f_0-a_0$ basis. Instead of ~(\ref{green4}) and ~(\ref{green5}), one gets
\be
\sum_{m=f,a} \left [ (E-E_n + i \frac{\Gamma_n}{2}) \delta_{nm} - \sum_{K \bar{K}} V_{n K \bar{K}} \frac{1}{E -
E_{K \bar{K}} + i \delta} V_{K \bar{K} m} \right ] G_{mk} = i \delta_{nk} ,
\label{green6}
\ee
where $n, k = f, a.$  The $a_0$-meson entering into this equation has been also ``renormalized,'' 
this time the
$\pi \eta$ channel plays the role of the $\pi \pi$ channel for $f_0$. 

Our next task is to present explicit
expressions for the sums over $K \bar{K}$ entering
 into~(\ref{green5}) and~(\ref{green6}).  This problem was
considered in~\cite{Bashinsky} in detail including the Coulomb interaction
 in the $K^+ K^-$ channel.  The Coulomb
interaction results in spectacular effects including the formation of the
 $K^+ K^-$ atom, but the typical energy
scale for these phenomena is of the order of few KeV
 which requires an experimental resolution that is probably
inaccessible in the near future. 
 Neglecting the  Coulomb interaction,  the above sums are reduced to 
expressions of the type
$$\sum_{K \bar{K}} V_{n K \bar{K}} Y (m_x, m_K) V_{K \bar{K} m}\ ,$$ where we replaced $E$ by $m_x$ to be
consistent with expressions~(\ref{cross2}-\ref{cross3}) for the cross sections; here,
 $Y (m_x, m_K)$ are  integrals of
the type
 
\be
Y (m_x, m_K) = \frac{1}{2m^3_x}\ \int \frac{d^3p}{(2 \pi)^3} \frac{1}{m_x - 2m_K - \frac{p^2}{m_K} + i \delta}   \ 
\ \ ,
\label{green7}
\ee
with $m_K = m_{K^{\pm}}, m_{K^{0}}$.  As already mentioned,
 the integral ~(\ref{green7}) is formally
 divergent which is however neither dangerous or important as soon as we focus
on the $m_x$ region
close to the $K \bar{K}$ threshold. 
 Close to the $K \bar{K}$ threshold the leading energy dependent
contribution is 

\be
 Y(m_x,m_K)=  
    -i \frac{1}{32\pi m_x}\ \sqrt{\frac{ m_x - 2m_K }{m_K} + i0 }   \  ,  
\label{Yterm}\ee where $+i0$ means that for
$m_x < 2\, m_K$ the square root acquires a positive
imaginary part.  From (~\ref{green6}) and (~\ref{Yterm}), 
it follows that for $m_x>2\, m_K$ the decay width into
the $K \bar{K}$ channel is given by $(n= f,a)$
\be
\Gamma_{n K \bar{K}} = \frac{\mid V_{n K \bar{K}} \mid^2}{ 16 \pi m_x} \sqrt{\frac{m_x - 2m_K }{m_K} }\ .
\label{green8a}
\ee  Comparison of (~\ref{green8a}) with the parameterization used in~\cite{Achasov1} shows that
$V_{n K \bar{K}} = g_{n K \bar{K}}, $ where $g_{n K \bar{K}}$ are the
coupling constants used in~\cite{Achasov1}.
 Recall that 
Eq.~(\ref{green6}) is
written in the basis of $f_0$ and $a_0$ states having
definite isospins, $I_{f_{0}} = 1, I_{a_{0}} = 0.$ 
Therefore,
\be
V_{K^+ K^- f} =V_{K^0 \bar{K}^0 f}\ \ \, \ \ 
 V_{K^+ K^- a} = -V_{K^0 \bar{K}^0 a}\ .
\label{green8}
\ee  Next we introduce the notation
\be
{\cal D} \equiv \frac{\mid  V_{K^+ K^- f} \mid^2 } {32 \pi m_x}\   \ , \ \
  \zeta \equiv \frac{ V_{K^+ K^- a}}{V_{K^+ K^- f} } \ \  ,
\label{green9}
\ee where $\zeta$ can be complex with phase $\phi,$ see later. 
In this notation, the explicit form of Eq.~(\ref{green6}) can be expressed
as a matrix in $f_0-a_0$ space as:
\begin{eqnarray}
i \hat{G}^{-1} =
\bigg( 
\begin{array}{cc}
m_x -E_f +i \frac{\Gamma_f}{2} & 0 \cr
 0 & m_x -E_a +i \frac{\Gamma_a}{2}  
\end{array}
\bigg) & & \nonumber \\
 + i {\cal D} 
\bigg( 
\begin{array}{cc}
1 & \zeta \cr
 \zeta^* & |\zeta|^2 
\end{array}) \bigg)
 \sqrt{ \frac{m_x-2 m_{K^+}}{m_{K^+} } + i0 }  
 & & \nonumber \\
 + i {\cal D} 
\bigg( 
\begin{array}{cc}
1 & -\zeta \cr
 -\zeta^* & |\zeta|^2 
\end{array} \bigg)
 \sqrt{ \frac{m_x-2 m_{K^0}}{m_{K^0} } + i0 }  
 \ \ ,
\label{green10}
\end{eqnarray}
Here we see that the isospin violating $f_0-a_0$ mixing
is really proportional to
$  \sqrt{ \frac{m_x - 2 m_K}{m_K}}$  and is indeed enhanced as was pointed
out in Ref.~\cite{Achasov2}.
We also see that far beyond the region $|m_x -2 m_K| \approx 8 MeV$ of the
$K\bar{K}$ thresholds, the
nondiagonal elements cancel each other and thereby extinguish the $f_0 - a_0$ mixing.  
Away from the threshold region one should also take into account
corrections to (~\ref{Yterm}), i.e., include the next terms 
in the expression for the renormalized kaon loop.

Now we have at our disposal
all quantities needed to calculate the near threshold
 amplitudes and cross sections
 using equations presented in Sec.~2. 
For example, the amplitude for the process
 $\gamma + p \rightarrow p' + f_0 \rightarrow p' + \pi \pi$ reads
\be
T_{\pi \pi} = V_{\pi \pi f} G_{ff} F_f + V_{\pi \pi f} G_{fa} F_a + B_{\pi \pi}.
\label{amppi}
\ee  Here we see the doorway production of the $f_0$, followed by
its subsequent propagation via $G_{ff}$ and then its decay $V_{f \pi \pi}.$
The second term includes the doorway production
of the $a_0,$ followed by its transition into a $f_0$ via isospin
violation and then the $f_0$ decays into a pion pair.  Other pion pair
processes are included in the background term $B_{\pi \pi}.$  For final
kaon production the corresponding result is:
\be
T_{K^+ K^-} = V_{K^+ K^- f}\ G_{ff}\ F_f + V_{K^+ K^- a}\ G_{aa}\ F_a +
V_{K^+ K^- f}\ G_{fa}\ F_a + V_{K^+ K^- a}\ G_{af}\ F_f + B_{K^+ K^-}\ \ .
\label{ampk} 
\ee
For final
$\pi \eta$ production the corresponding result is:
\be
T_{\pi \eta} = V_{\pi \eta a}\ G_{aa}\ F_a + V_{\pi \eta a}\ G_{af}\ F_f + B_{\pi \eta}.
\label{amppe} 
\ee
The values of the parameters are discussed in the next section.

\section{Parameters of the Model}

The  theoretical investigation of the light scalar mesons $f_0$ and $a_0$ has
a long history. A most thoughtful study of $f_0/a_0$ has been
performed by the Novosibirsk group (see~\cite{Achasov1,Novo} and references
therein). Although our approach differs from other authors,
including the Novosibirsk group, the key parameters are similar in
various treatments. Thus, in order to fix the parameters we
shall relate them to those used by the Novosibirsk group.  This would still leave
the set of parameters incomplete, namely the photoproduction amplitude $F_k$ and 
the background term
$B_\alpha$ ( see Eq.\ref{amp2}) have no analogues in the Novosibirsk set and
 to get at least an educated guess
about them, we resort to the recent paper~\cite{MOT}.  We stress
that the characteristics of
$f_0/a_0$ are at present far from being perfectly established.  For example,
the Novosibirsk group presents several solutions with
some of the parameters ranging over rather wide limits~\cite{Achasov1}.
In order to determine the parameters of the model we use the
recent experimental data presented in~\cite{Novo}. These data do not 
allow us to
fix all the parameters and some parameters will be set from different
consideration (see below).

The model under consideration can be described by several equivalent
sets of parameters. We used the following set: $E_f,E_a,\Gamma_f,
\Gamma_a, V_{f K^+ K^-}$, and $\zeta$. The quantities
$V_{f K^0\bar{K}^0}$, $V_{aK^0\bar{K}^0}$ and ${\cal D}$ are then
expressed according to (~\ref{green8}) and (~\ref{green9}),
 while $V_{f \pi\pi}$ and
$V_{a\pi \eta}$ are determined via $\Gamma_f$ and $\Gamma_a$ according
to the equation

\begin{equation}
\Gamma_{j m_1 m_2} = \frac{|V_{j m_1 m_2}|^2}{16\pi m_x}
\rho_{m_1 m_2}\, ,
\label{5.1}
\end{equation}
where $j = f,a$, and $\rho_{m_1 m_2}$ is the two-body phase space.
At this point we remind the reader
 that since we treat the $K \bar{K}$ channels
explicitly, their contribution does not enter into $E_j$ and
$\Gamma_j (j = f,a)$. We also mention that the ``visible" widths of
$f_0/a_0$ are smaller
 than the widths $\Gamma_f$ and $\Gamma_a$~\cite{Achasov1}.
Finally, we recall that our parameters $V_{j m_1 m_2}$ are equivalent
to $g_{j m_1 m_2}$ used by the Novosibirsk
 group~\cite{Achasov1}~\cite{Novo}.

In~\cite{Novo} we find the following parameters relevant for our purposes
(experimental errors are omitted):

\begin{equation}
\frac{|V_{f \pi \pi}|^2}{4\pi} = 0.44\  GeV^2\;,
R = \frac{|V_{f K\bar{K}}|^2}{|V_{f \pi \pi}|^2} = 3.77\,.
\label{5.2}
\end{equation}
Using (~\ref{5.1}) and isotopic relations
\begin{equation}
|V_{f \pi^+ \pi^-}|^2 = \frac{2}{3} |V_{f \pi \pi}|^2,\ \ \ \ 
|V_{f K^+ K^-}|^2 = \frac{1}{2} |V_{f K \bar{K}}|^2,
\label{5.3}
\end{equation}
we get from (~\ref{5.2}):
\begin{equation}
\Gamma_f = 0.1\ GeV\,, V_{f K^+ K^-} =3.23\ GeV\,.
\label{5.4}
\end{equation}
Next setting in (~\ref{green10}) $m_x$ equal to the physical mass of the
$f_0$-meson, i. e. $m_x = 0.98$~GeV, we find $E_f = 0.947$~GeV.
Unfortunately, we can not rely on well-established experimental data for
the parameters of the $a_0$ meson.
Therefore we assume $\Gamma_a \equiv \Gamma_f$ and $E_a \equiv E_f$ (these
assumptions are in line with the solutions proposed in~\cite{Achasov1}.)
 From
$\Gamma_a$ we get $V_{a \pi \eta} = 3.08$~GeV. Finally we set
$\zeta = 1,$ which is true in both four-quark and molecular models of
$f_0/a_0$~\cite{Achasov1}. Thus we arrive at the following parameter set (set A):

\begin{equation}
E_f = E_a = 0.947\ GeV, \; \Gamma_a = \Gamma_f = 0.1\ GeV, \; \\
V_{f K^+ K^-} = 3.23\ GeV, \;  \zeta =1.
\label{5.5}
\end{equation}
Note that the value $|V_{f K^+ K^-}|^2/4\pi = 0.83$~GeV$^2$ lies in
between the values typical for four-quark and molecular models of
$f_0$-meson~\cite{Achasov1}.

In order to display the effect of $f_0-a_0$ mixing, we consider also
set B whose only difference from set A is that
 in B the splitting of
the $K^0 \bar{K}^0$ and
$K^+ K^-$ thresholds
is ignored and the kaon mass is taken equal to the average value
$m_K = \frac{1}{2} ( m_{K^0} + m_{K^\pm} )$ and hence mixing is thereby turned off.
This results in cancellation of the non-diagonal terms in Eq.~\ref{green10}.

Within our approach we are not in a position
to determine the value of the
photoproduction amplitude $F_j (j = f,a).$
 Its invariant structure is given by Eqs.~\ref{invM} and  ~\ref{inv2}, 
but the magnitudes of the invariant amplitudes remain unknown.
In a recent paper~\cite{MOT} the
$f_0/a_0$ photoproduction has been
considered in the near-threshold region, namely at $k= 1.7\ 
 GeV.$  Close
to threshold the dominant contribution stems from the
amplitude $R_1$ of Eq.~\ref{inv2}.  Its value can be considered a free
parameter, but  we prefer
  to take for $R_1$ the value 
proposed in~\cite{MOT}.~\footnote{This threshold value for $R_1$ is basically
the pseudoscalar photoproduction Kroll-Ruderman limit scaled by a factor $k/(2 M)$
due to the switch to a scalar meson}  The reasons for adopting
 this value for $R_1$ is
on one hand because it enables a direct comparison of our
results with that of~\cite{MOT} and on the other hand, as 
already argued in Sec.~2,  we may consider the pion loops
included into the renormalized vertex of $f_0$ photoproduction
while in the approach of~\cite{MOT}
 the two pion photoproduction is the
 driving mechanism that determines the value of $R_1.$

The calculation or even estimate of the
background term $B_\alpha$ in ~\ref{amp1} is
beyond the scope of this article.  Some of the diagrams entering
into $B_\alpha$ are depicted in Fig.~2 and in general may be
calculated.  In Ref.~\cite{MOT}
the background for $d \sigma_{\pi^+ \pi^-}/d m_x$ (see ~\ref{cross3})
was estimated to be around $55\ \mu b/GeV.$

In Fig. 4,  we present the invariant mass distribution,
$d \sigma(K^+ K^-)/d m_x ,$
 for $K^+ K^-$ pair
production calculated according to Eqs. 2.5, 4.17.and 4.19.
 In order to display the role
of isospin breaking $f_0-a_0$ mixing, 
we plot on the same figure the curve obtained with set B parameters, 
i.e., with $f_0-a_0$
mixing switched off (which is achieved by using the same mass for the
charged and neutral kaons).  The two curves are different, in line
with ``enhanced" mixing in the sense described above. (The scalar meson mixing
can be as large as a 70 \% enhancement.) The
structure in the invariant mass distribution for $K^+ K^-$ pair
production 
in this near threshold region arises from the
isospin breaking difference in mass between charged and neutral kaons. 

The $a_0-f_0$ interference pattern turns out to
be very sensitive to the phase of the parameter
$\zeta$ (see (4.16)).  The sensitivity to that phase is
clearly displayed in Fig.~4.  One should however keep in mind
that in both $q \bar{q}$ and $q^2 \bar{q}^2$ constituent quark
models this parameter is predicted to be
real~\cite{Novo,Jaffe}.

In Fig.5, the $\pi^+ \pi^-$ invariant mass distribution
$d \sigma(\pi^+ \pi^-)/d m_x$ is presented.
We now use Eqs. 2.5, 4.17.and 4.18. 
Here the deviation from a simple Breit-Wigner resonance structure due
to the influence of the
$K \bar{K}$ thresholds is clearly seen.
We do not insist on the absolute values of the cross sections
 in Figs. 4 and 5,  because of
backgound effects,  but we do believe the change due to mixing
is of some predictive value.

\section{Conclusions}

The analysis presented herein illustrates that threshold photoproduction 
can provide insight into the nature of
the $f_0-$ and $a_0$ scalar mesons.  Valuable information may
be obtained from the
$K \bar{K}$ mass distribution in the threshold region if 
measured with a few MeV resolution.  In our approach, this distribution
has been described in a model-independent way, once the doorway idea is invoked. Most
parameters can be deduced from $\phi \rightarrow \gamma (a_0 + f_0)
\rightarrow K \bar{K}$ experiments.

The invariant amplitude formalism for the photoproduction
of scalar mesons presented in Sec. III, enables us to consider higher energy
regions which are accessible at JLAB.  Going to higher energies increases the
number of CGLN amplitude parameters by involving all of the $R_i$ terms
in (3.4).  Also, with increasing energy more partial waves come into play 
giving rise to nonzero spin observables.  The role of spin observables
and how they evolve with increasing energy will be discussed in
a future paper.

\acknowledgments
We wish to thank Dr. S. Bashinsky and Profs. W. Kloet, J. Thompson,
S. Eidelman,
and  E.P. Solodov for their helpful suggestions.   
We also thank Professor R.~Schumacher for information
about scalar mesons.
B.K. wishes to express appreciation for 
warm hospitality and financial support during his visit  to
the University of Pittsburgh and for support by RFFI grants 97-02-16406
and 00-02-17836, 
as well as from the U.S. National Science Foundation.


%
%

\newpage
\begin{figure}
\epsfysize= 4in
\epsfbox{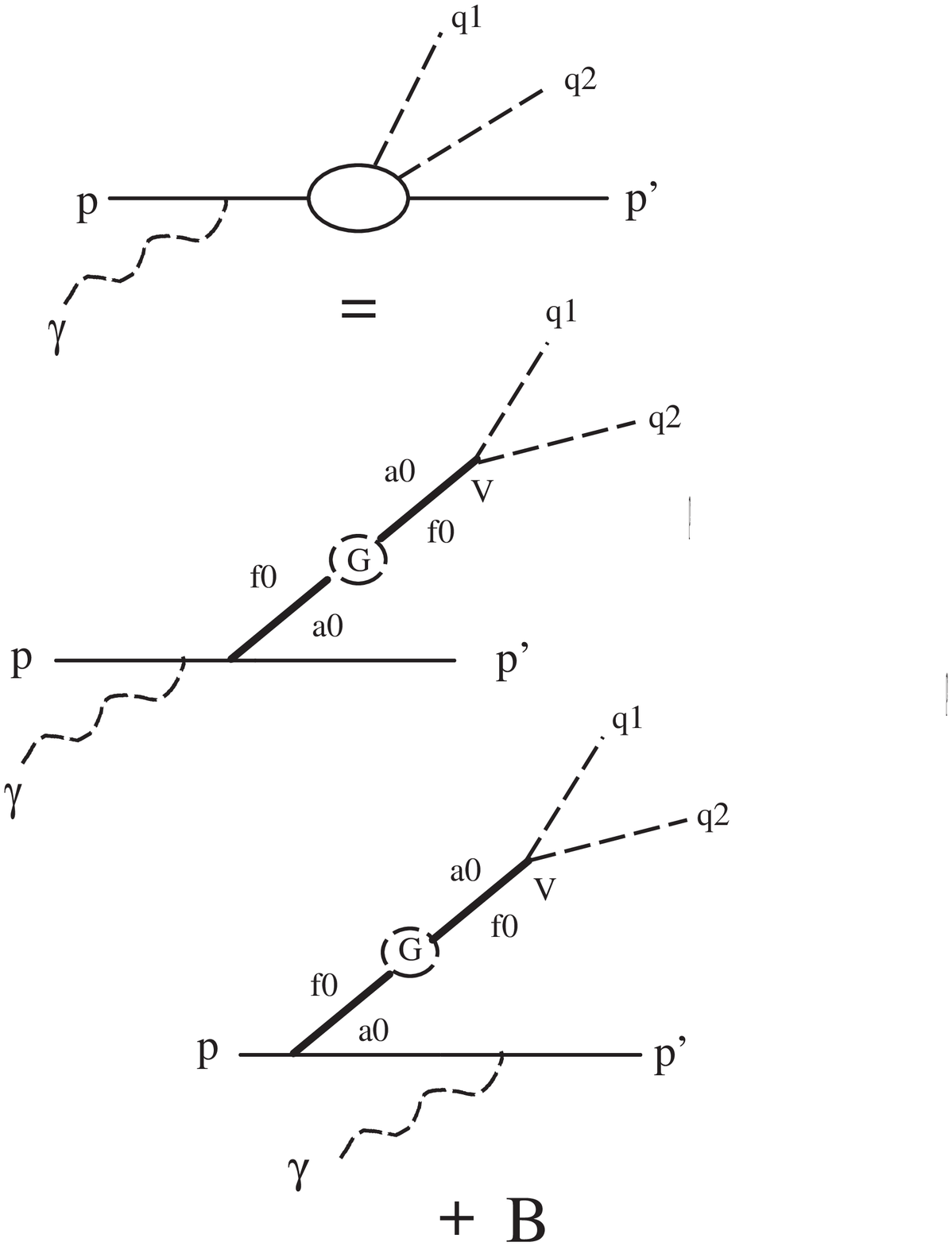}
\vskip 1in
\caption{ Feynman diagrams describing the doorway production of the 
reaction $\gamma p\to p' m\bar
m$.  The total amplitude is a sum of these $f_0/a_0$  contributions plus
the background $B,$ which is described in Fig. 2. The  solid line with the blob $G$
 insertion denotes 
the complete $f_0/a_0$ propagator (see Eq.~\ref{green10}) with $f_0 \leftrightarrow a_0$
mixing. The decay mesons shown as dashed lines have momenta $q_1$ and $q_2.$ 
} 
\end{figure}
%
\begin{figure}
\epsfysize= 4in
\epsfbox{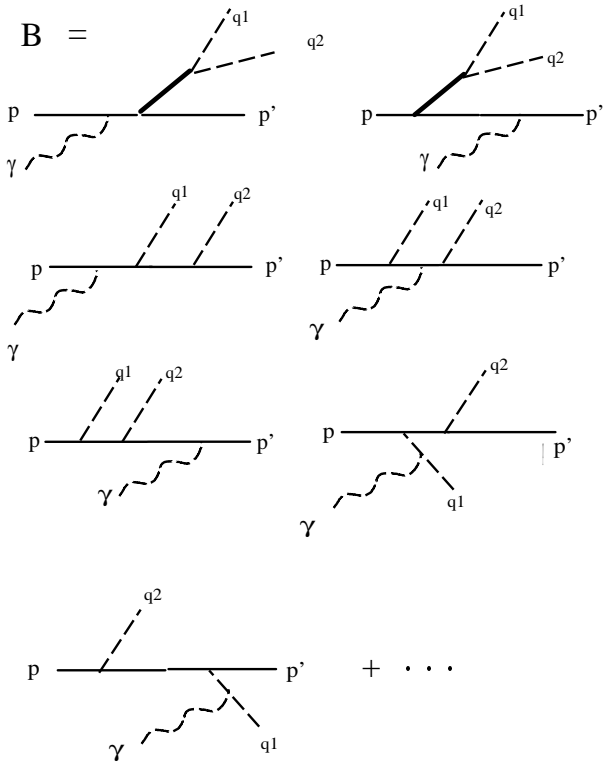}  
\caption{ Feynman diagrams describing the possible background contributions
 $B$ for the
 the reaction $\gamma p\to p' m\bar m.$  
 Dark solid lines stand for the contribution of
resonances other than $f_0/a_0.$}  
\end{figure}
%
\begin{figure}
\epsfysize= 5in
\epsfbox{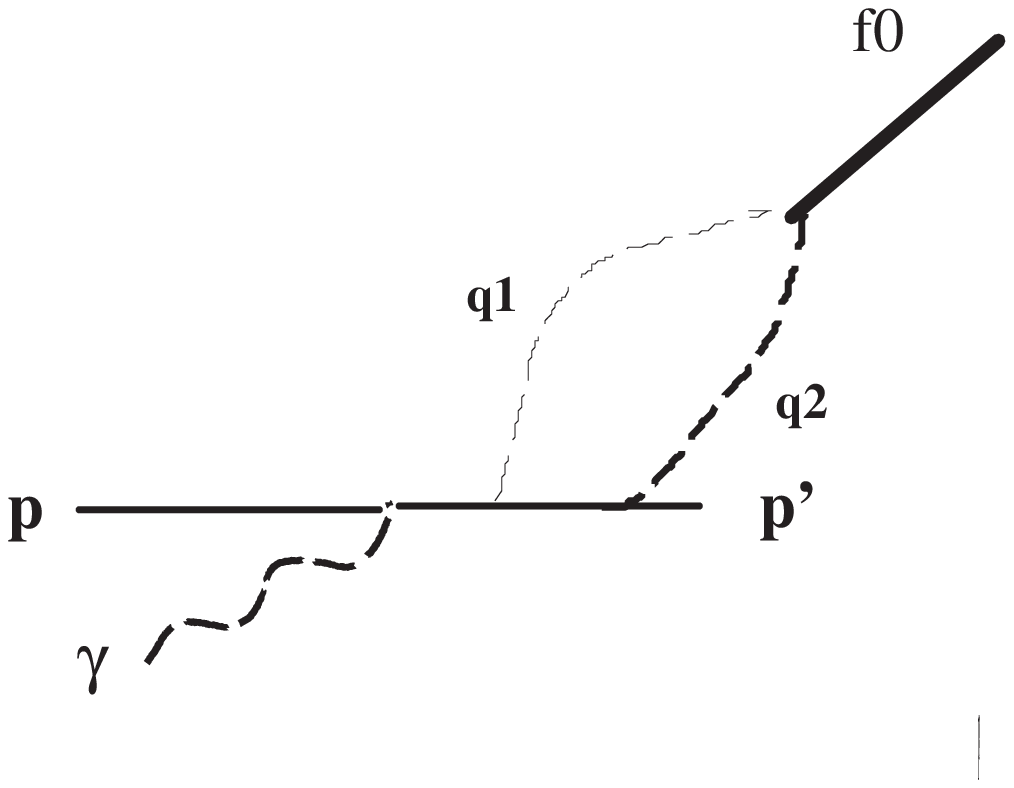}
\vskip -3in
\caption{ Production of $f_0$ via the pion loop as discussed in the text.}
\end{figure}
%
\begin{figure}
\epsfysize= 6in
\epsfbox{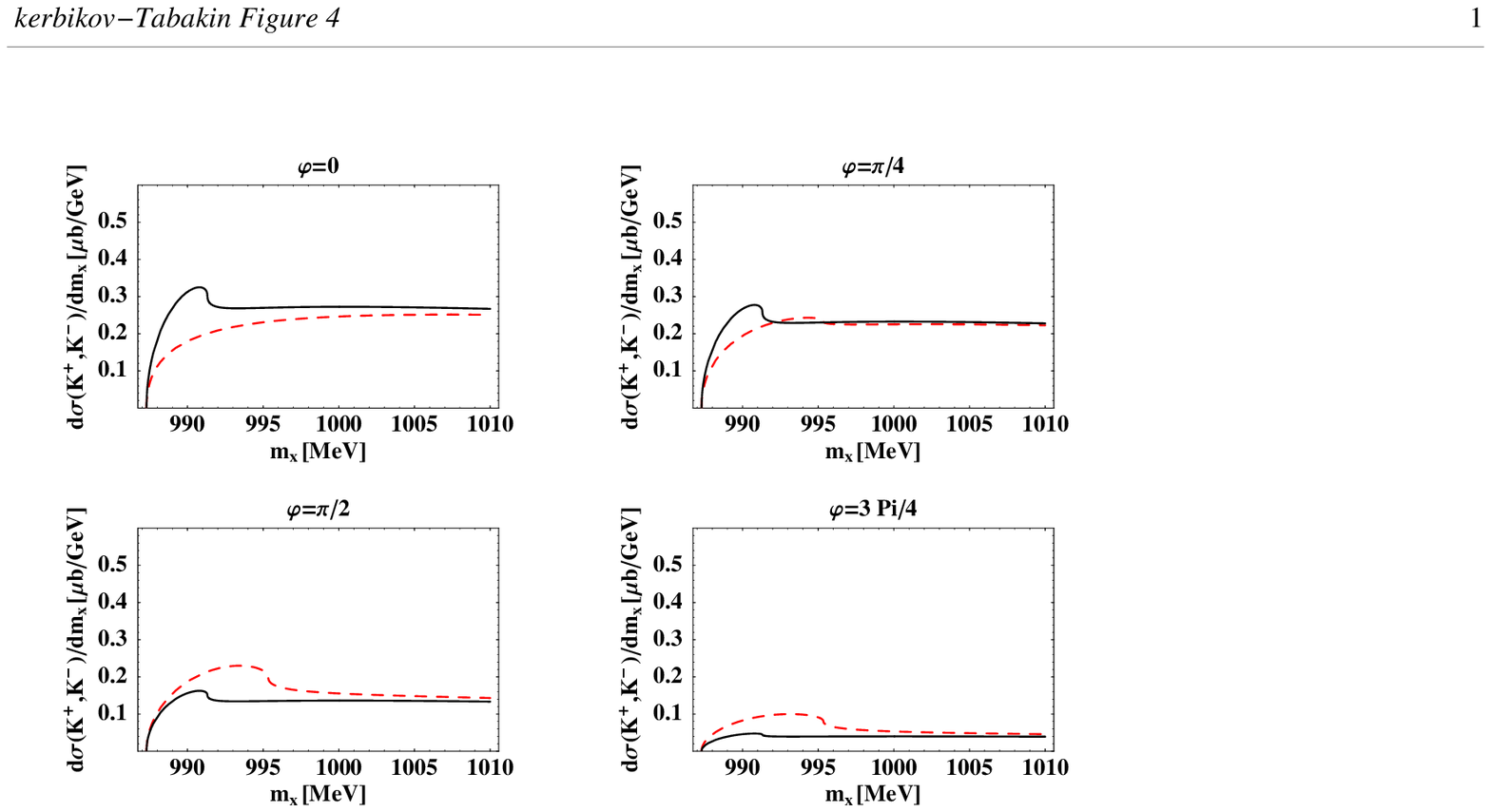}
\vskip 1in
\caption{ 
The invariant mass distribution
$d \sigma(K^+ K^-)/d m_x$ in microbarns/GeV is plotted versus
the invariant mass of the 
  $K^+ K^-$ pair resulting from scalar meson photoproduction.
The dashed curve includes the $f_0 \leftrightarrow a_0$ transitions
caused by isospin breaking, whereas the solid curve is for case B
 for which the charged and neutral kaon masses are set equal
 to their average and the
isospin effect is thus turned off. 
 The dependence on the phase of the ratio
 $\zeta = |\zeta|\exp{i \phi}  \equiv  V_{K^+ K^- a}/V_{K^+ K^- f } $
is also shown, including the associated phase on the decay amplitudes
 $V_{K^+ K^- a}$ and $V_{K^+ K^- f }.$
}
\end{figure}
%
\begin{figure}
\epsfysize= 4in
\epsfbox{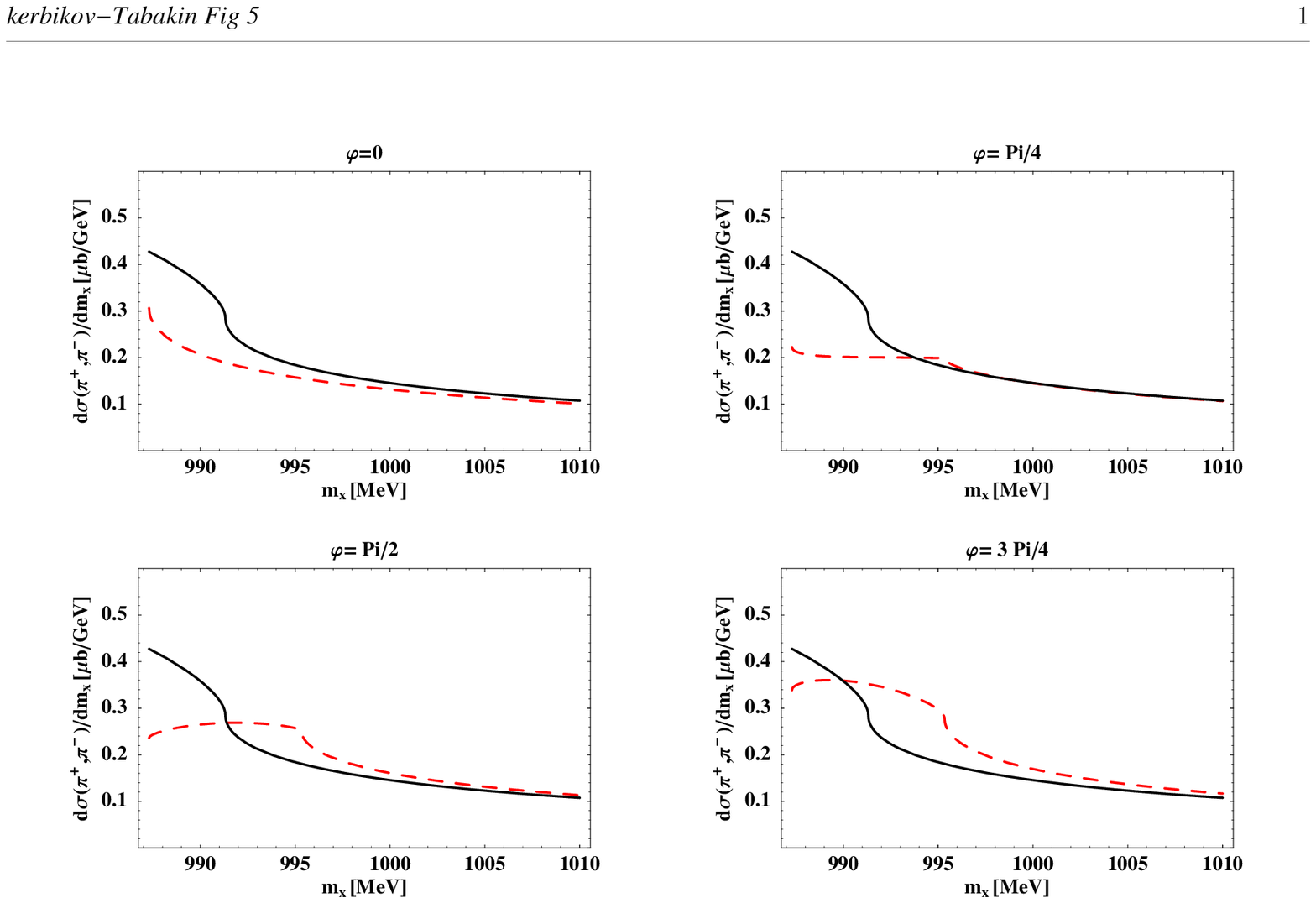}
\vskip 1in
\caption{  
The invariant mass distribution
$d \sigma(\pi^+ \pi^-)/d m_x$ in microbarns/GeV is plotted versus
the invariant mass of the 
  $\pi^+ \pi^-$ pair resulting from scalar meson photoproduction.
The dashed curve includes the $f_0 \leftrightarrow a_0$ transitions
caused by isospin breaking, whereas the solid curve is for set B
 for which the charged and neutral kaon masses are set equal
 to their average and the
isospin effect is thus turned off. 
 The dependence on the phase of the ratio
 $\zeta = |\zeta|\exp{i \phi}  \equiv  V_{K^+ K^- a}/V_{K^+ K^- f} $
is also shown. Here 
 $V_{\pi^+ \pi^- f} $ is taken as real. 
 The same value of $R_1$ is used as in the kaon pair case.
}
\end{figure}
\end{document}